# A New Approach to the Problem of Dynamical Quarks in Numerical Simulations of Lattice QCD

Martin Lüscher

Deutsches Elektronen-Synchrotron DESY
Notkestrasse 85, D-22603 Hamburg, Germany

## Abstract

Lattice QCD with an even number of degenerate quark flavours is shown to be a limit of a local bosonic field theory. The action of the bosonic theory is real and bounded from below so that standard simulation algorithms can be expected to apply. The feasibility of such calculations is discussed, but no practical tests have yet been made.

## 1. Introduction

It is well-known that the quark degrees of freedom are difficult to include in numerical simulations of lattice QCD. At present the probably best method is to integrate over the quark fields analytically and to simulate the resulting effective gauge field theory using the Hybrid Monte Carlo algorithm [1]. This ingenious algorithm represents a significant progress compared to earlier techniques, but the computer resources required for QCD simulations remain enormous (see ref.[2], for example).

In this paper a different computational strategy is proposed. No practical tests of the algorithm have yet been performed and much further work will be needed before the merits and limitations of the method can be assessed. The basic idea is to map the lattice theory to a local bosonic theory by incorporating a suitably chosen fermion matrix inversion algorithm in the functional integral. Simulation methods familiar from pure gauge theories and Higgs models may then be applied. In particular, all degrees of freedom are treated stochastically and no exact inversion of the Dirac operator is required in the course of the simulation.

The situation is, however, not quite so simple, because the bosonic theory involves a large number of complex fields coupled to the gauge field. One actually needs to take the number of fields to infinity if lattice QCD is to be reproduced exactly. To avoid excessive memory requirements, the bosonic theory should hence be constructed in such a way that the limit is reached rapidly. A concrete proposal on how to accelerate the convergence is included in the present paper, and a number of further technical questions concerning the simulation of the bosonic theory are addressed.

## 2. Lattice QCD

In the following attention is restricted to the case of two degenerate flavours of Wilson quarks. Improved Wilson fermions or staggered fermions do not present any additional difficulties.

We are thus concerned with gauge fields $U(x,\mu)$ and quark fields $\psi(x)$, $\overline{\psi}(x)$ residing on a 4-dimensional hyper-cubic lattice with spacing $a$ and vertices $x$. The link variables $U(x,\mu)$ take values in SU(3) and the fermion fields carry flavour, colour and Dirac indices. The lattice is assumed to be finite with periodic or anti-periodic boundary conditions (boundary conditions as required for the QCD Schrödinger functional [6] are also tolerated).

The action of the theory is

$$S[U,\overline{\psi},\psi] = S_g[U] + a^4 \sum_x \overline{\psi}(x)(D+m)\psi(x), \tag{2.1}$$

where $S_g[U]$ denotes the Wilson plaquette action, $m$ the bare quark mass and

$$D = \tfrac{1}{2} \sum_{\mu=0}^{3} \{\gamma_\mu (\nabla_\mu^* + \nabla_\mu) - a\nabla_\mu^* \nabla_\mu\} \tag{2.2}$$

the lattice Dirac operator. Hermitean $\gamma$–matrices are assumed here and the covariant forward difference operator is given by

$$\nabla_\mu \psi(x) = \frac{1}{a}\left[U(x,\mu)\psi(x+a\hat{\mu}) - \psi(x)\right], \tag{2.3}$$

with $\hat{\mu}$ being the unit vector in the positive $\mu$ direction. The backward difference operator $\nabla_\mu^*$ is equal to minus the adjoint of $\nabla_\mu$.

After integrating over the quark fields, the effective gauge field distribution becomes

$$P_{\text{eff}}[U] = \frac{1}{\mathcal{Z}} \left[\det(D+m)\right]^2 \mathrm{e}^{-S_g[U]}. \tag{2.4}$$

The normalization constant $\mathcal{Z}$ (the partition function) is defined such that

$$\int \mathrm{D}[U]\, P_{\text{eff}}[U] = 1, \tag{2.5}$$

where one integrates over all gauge fields $U$ and $\mathrm{D}[U]$ denotes the usual SU(3) invariant product measure. Note that it is the square of the Dirac determinant which enters here, because two degenerate flavours of quarks have been assumed. Our aim in sect. 3 will be to find an exact representation of $P_{\text{eff}}[U]$ in terms of a local bosonic field theory.

The lattice Dirac operator is a sum of an anti-hermitean and a hermitean part. If we define $\gamma_5 = \gamma_0 \gamma_1 \gamma_2 \gamma_3$, it is however easy to show that

$$[\gamma_5(D+m)]^\dagger = \gamma_5(D+m). \tag{2.6}$$

The operator $\gamma_5(D+m)$ thus has a complete set of eigenvectors with real eigenvalues. In particular, the quark determinant $\det(D+m)$ is real and the distribution (2.4) defines a probability measure.

Without loss of generality the quark mass $m$ may be chosen such that the hopping parameter $\kappa = (8+2am)^{-1}$ is non-negative. One may then prove that

$$\|\gamma_5(D+m)\| \leq M, \qquad M = 8/a + m. \tag{2.7}$$

In the following it will be convenient to work with the normalized quark matrix

$$Q = \gamma_5(D+m)/M, \tag{2.8}$$

which has eigenvalues between $-1$ and $1$.

## 3. Transformation to a local bosonic theory

In the first step of the transformation we need to choose a polynomial $P(s)$ which approximates the function $1/s$ in the interval $0 \leq s \leq 1$. For example, we may take

$$P(s) = \sum_{k=0}^{n} (1-s)^k, \tag{3.1}$$

which obviously satisfies

$$\lim_{n \to \infty} P(s) = 1/s \quad \text{for} \quad 0 \leq s \leq 1. \tag{3.2}$$

This particular polynomial is not the best choice, but to explain the transformation it is helpful to start with a simple case. A better polynomial will be discussed in sect. 4.

Since all eigenvalues of $Q^2$ are between 0 and 1, the matrix $P(Q^2)$ converges to the propagator $1/Q^2$. In particular,

$$\det Q^2 = \lim_{n \to \infty} \left[\det P(Q^2)\right]^{-1}. \tag{3.3}$$

Note, incidentally, that any polynomial approximation $P(s)$ to the function $1/s$ defines an algorithm to solve the linear system $Q^2\psi = \eta$. One simply evaluates $P(Q^2)\eta$ using some recursive scheme as in refs.[3–5].

The roots $z_k$, $k = 1, \ldots, n$, of the polynomial (3.1) are given by

$$z_k = 1 - \exp\left\{i\frac{2\pi k}{n+1}\right\}. \tag{3.4}$$

For even $n$ they come in complex conjugate pairs with non-zero imaginary parts. The polynomial may then be written in the manifestly positive form

$$P(Q^2) = \prod_{k=1}^{n}\left[(Q - \mu_k)^2 + \nu_k^2\right], \tag{3.5}$$

where the $\mu_k$'s and $\nu_k$'s are determined through

$$\mu_k + i\nu_k = \sqrt{z_k}, \qquad \nu_k > 0. \tag{3.6}$$

Every value of $\nu_k$ occurs twice and the corresponding $\mu_k$'s have equal magnitude and opposite sign.

The bosonic theory alluded to above couples the gauge field to $n$ complex fields $\phi_k(x)$ with colour and Dirac indices but no flavour index ($n$ is taken to be even in the following). The action is

$$S_b[U, \phi] = S_g[U] + a^4 \sum_x \sum_{k=1}^{n}\left\{\left|(Q - \mu_k)\phi_k(x)\right|^2 + \nu_k^2 \left|\phi_k(x)\right|^2\right\}, \tag{3.7}$$

and from the above we now infer that

$$P_{\text{eff}}[U] = \lim_{n \to \infty} \frac{1}{\mathcal{Z}_b} \int \mathrm{D}[\phi]\mathrm{D}[\phi^\dagger]\, \mathrm{e}^{-S_b[U,\phi]}, \tag{3.8}$$

where $\mathcal{Z}_b$ denotes the partition function of the bosonic system,

$$\mathcal{Z}_b = \int \mathrm{D}[U]\mathrm{D}[\phi]\mathrm{D}[\phi^\dagger]\, \mathrm{e}^{-S_b[U,\phi]}. \tag{3.9}$$

It should be emphasized that the action $S_b[U, \phi]$ is local and non-negative. In particular, the Gaussian integrals in eq.(3.8) are well-defined.

The construction of the bosonic theory can be carried out in exactly the same way for almost any polynomial approximation $P(s)$ to the function $1/s$. One only requires that eq.(3.2) holds and that none of the roots $z_k$ of the polynomial are real.

## 4. Chebyshev acceleration

For the numerical simulation of the bosonic theory, $n$ is set to some large but finite value, depending on the lattice size and the physical conditions. One must then make sure that the systematic error arising from the finiteness of $n$ is negligible compared to the statistical errors. Since one cannot afford to take $n$ to arbitrarily large values, it is important to achieve rapid convergence by optimizing the polynomial $P(s)$.

A similiar optimization problem occurs when solving systems of linear equations through an iterative procedure (see chapt. 4 of ref.[5] for a particularly readable account). It is well-known that the convergence of some of the algorithms can be significantly accelerated using Chebyshev polynomials [3–5]. While the situation here is slightly different, the basic ideas carry over and lead to the optimized polynomial $P(s)$ defined below.

*4.1 Convergence criteria*

The rate of convergence of a matrix inversion algorithm depends on the condition number $p$ of the matrix. For a positive hermitean matrix, $p$ is equal to the ratio of the largest to the smallest eigenvalue. The condition number of the fermion matrix $Q^2$ is roughly proportional to the square of the inverse lattice spacing and is hence rather large in the cases of interest.

The rate of convergence of the algorithm derived from the polynomial (3.1) can be determined as follows. First note that

$$|P(s) - 1/s| = (1-s)^{n+1}/s \quad \text{for} \quad 0 \leq s \leq 1. \tag{4.1}$$

Exponential convergence is thus guaranteed for all spectral values $s > 0$, but the rate is rapidly varying and goes to zero when $s$ becomes small. If we assume that the largest eigenvalue of the quark matrix $Q^2$ is close to 1 (which is plausible since $Q$ has been normalized), the condition number $p$ will be approximately equal to the inverse of the smallest eigenvalue of $Q^2$. We then conclude that $P(Q^2)$ converges to $1/Q^2$ exponentially with a rate close to $1/p$.

Good inversion algorithms achieve much better rates, equal to $4/\sqrt{p}$ or at least $2/\sqrt{p}$ [5]. Our aim in the following will be to construct a polynomial related to such an algorithm.

A further criterion for a good choice of polynomial is obtained when considering the force acting on the gauge field. The force is proportional to the variation $\delta S_{\text{eff}}[U]$ of the action of the effective gauge theory under small changes of the link variables. After replacing $\det Q^2$ by $[\det P(Q^2)]^{-1}$, the variation of the action is given by

$$\delta S_{\text{eff}}[U] = \delta S_g[U] + \text{Tr}\left\{\delta Q^2 P'(Q^2)/P(Q^2)\right\}, \tag{4.2}$$

where $P'(s)$ denotes the derivative of $P(s)$. The force, or some integrated form of it, drives the stochastic evolution of the gauge field in numerical simulations of QCD. We should, therefore, also make sure that the ratio $P'(s)/P(s)$ converges rapidly to $-1/s$.

*4.2 Chebyshev polynomials*

For any real number $u$ between 0 and 1 an angle $\theta$ may be defined through

$$\cos\theta = 2u - 1, \qquad 0 \leq \theta \leq \pi. \tag{4.3}$$

The (modified) Chebyshev polynomial $T_r^*(u)$ of degree $r$ is then given by

$$T_r^*(u) = \cos(r\theta). \tag{4.4}$$

In particular,

$$T_0^*(u) = 1, \qquad T_1^*(u) = 2u - 1, \tag{4.5}$$

and the higher-order polynomials may be worked out recursively using

$$uT_r^*(u) = \tfrac{1}{4}\left\{T_{r+1}^*(u) + T_{r-1}^*(u) + 2T_r^*(u)\right\}, \qquad r \geq 1. \tag{4.6}$$

From the definition (4.4) it is obvious that

$$\sup_{0 \leq u \leq 1} |T_r^*(u)| = 1, \tag{4.7}$$

and the relation

$$\frac{\mathrm{d}}{\mathrm{d}u}\left\{\frac{T_{r+1}^*(u)}{r+1} - \frac{T_{r-1}^*(u)}{r-1}\right\} = 4T_r^*(u) \tag{4.8}$$

is also obtained with little effort.

Chebyshev polynomials are uniformly bounded in the interval $0 \leq u \leq 1$. Outside this range they are rapidly and monotonically increasing in magnitude. For $u < 0$, for example, we have

$$T_r^*(u) = (-1)^r \cosh(r\chi), \tag{4.9}$$

where $\chi > 0$ is determined through

$$\cosh \chi = 1 - 2u. \qquad (4.10)$$

4.3 Definition of $P(s)$

We first seek a polynomial approximation to the function $1/s$ in the interval $\varepsilon \leq s \leq 1$, where $\varepsilon$ is an adjustable parameter satisfying

$$0 < \varepsilon < 1. \qquad (4.11)$$

To this end the fit range is mapped to the interval $0 \leq u \leq 1$ through the transformation

$$s \to u = (s - \varepsilon)/(1 - \varepsilon). \qquad (4.12)$$

We then consider the polynomial

$$R(s) = \rho \left\{ \frac{T^*_{n+1}(u)}{n+1} - \frac{T^*_{n-1}(u)}{n-1} \right\} \qquad (4.13)$$

and choose the constant $\rho$ such that

$$R(0) = -1. \qquad (4.14)$$

As will become clear below, $R(s)$ plays the rôle of an error term and was chosen with care so as to fulfill the criteria discussed in subsect. 4.1. Explicitly one finds

$$\rho = \left\{ \frac{\cosh\bigl((n+1)\alpha\bigr)}{n+1} - \frac{\cosh\bigl((n-1)\alpha\bigr)}{n-1} \right\}^{-1}, \qquad (4.15)$$

where $\alpha > 0$ is determined by

$$\cosh \alpha = (1 + \varepsilon)/(1 - \varepsilon) \qquad (4.16)$$

(here and below we assume that $n$ is even).

We now define a polynomial $P(s)$ of degree $n$ through

$$P(s) = [1 + R(s)]/s. \qquad (4.17)$$

Note that $1 + R(s)$ vanishes at $s = 0$ and is therefore divisible by $s$. From the definition of $R(s)$ and the uniform boundedness of the Chebyshev polynomials it is clear that

$$|P(s) - 1/s| \leq \frac{2|\rho|}{(n-1)s} \quad \text{for} \quad \varepsilon \leq s \leq 1. \tag{4.18}$$

Furthermore, since the constant $\rho$ goes to zero exponentially at large $n$,

$$\rho = \frac{n}{\sinh \alpha} e^{-n\alpha} \{1 + \mathrm{O}(1/n)\}, \tag{4.19}$$

we conclude that $P(s)$ approximates $1/s$ in the range $\varepsilon \leq s \leq 1$, with an absolute relative error which is uniformly bounded and exponentially small.

The polynomial in fact converges to $1/s$ for all $s$ between 0 and 1, although with a gradually decreasing exponential rate when $0 \leq s < \varepsilon$. More precisely, the rate in this range is equal to $\alpha - \chi$, where $\chi$ is determined through eqs.(4.10) and (4.12). The maximal value,

$$P(0) = \frac{4\rho}{1-\varepsilon} \cosh(n\alpha) = \frac{2n}{(1-\varepsilon)\sinh \alpha} \{1 + \mathrm{O}(1/n)\}, \tag{4.20}$$

diverges linearly with $n$, with a large coefficient if $\varepsilon$ is small.

*4.4 Properties of $P'(s)$*

As discussed in subsect. 4.1, one also requires that $P'(s)/P(s)$ is a good approximation to the function $-1/s$ in the range $0 \leq s \leq 1$.

The particular form (4.13) of the polynomial $R(s)$ was chosen so that

$$R'(s) = \frac{4\rho}{1-\varepsilon} T_n^*(u) \tag{4.21}$$

[cf. eq.(4.8)]. The error term in the relation

$$P'(s)/P(s) = -1/s + R'(s)/[1 + R(s)] \tag{4.22}$$

(which one deduces straightforwardly from the definition of $P(s)$) is hence uniformly bounded in the range $\varepsilon \leq s \leq 1$. In particular, at large $n$ the bound

$$|P'(s)/P(s) + 1/s| \leq \frac{4n}{(1-\varepsilon)\sinh \alpha} e^{-n\alpha} \{1 + \mathrm{O}(1/n)\} \tag{4.23}$$

is obtained.

When $s$ is between $0$ and $\varepsilon$, the ratio $P'(s)/P(s)$ continues to converge to $-1/s$, with a reduced exponential rate equal to $\alpha - \chi$.

*4.5 Computation of the roots $z_k$*

To write down the action of the bosonic theory derived in sect. 3, one needs to know the roots $z_k$ of the polynomial $P(s)$. It does not seem possible to obtain them in closed analytic form. They can always be computed numerically, of course, to any desired precision. Since the degree $n$ can be quite large, it is however not advisable to use standard library routines for this task. A better way to proceed is described here, and we shall also obtain some useful qualitative information on the distribution of the roots in the complex plane.

Since we are mainly interested in the cases where $n$ is large, we may without loss assume that $0 < \rho < \frac{1}{4}$. The polynomial $P(s)$ can then be shown to be positive for all $s$ and its roots $z_k$ thus come in complex conjugate pairs with non-zero imaginary parts.

A systematic expansion of the roots in powers of $1/n$ may be deduced as follows. First note that the $n+1$ solutions of the equation

$$1 + R(s) = 0 \qquad (4.24)$$

are $s = 0$ and $s = z_k$, $k = 1, \ldots, n$. Now when $s$ has a non-zero imaginary part, there is a unique complex number $w$ such that

$$\tfrac{1}{2}(w + 1/w) = 2u - 1, \qquad |w| > 1. \qquad (4.25)$$

In terms of this variable, eq.(4.24) can be rewritten in the form

$$w^{n+1} = -r^{n+1} f(w), \qquad (4.26)$$

where $r > 0$ and $f(w)$ are defined through

$$r = \{2(n+1)/\rho\}^{1/(n+1)}, \qquad (4.27)$$

$$f(w) = \left\{1 - [(n+1)/(n-1)](w^{-2} + w^{-2n}) + w^{-2n-2}\right\}^{-1}. \qquad (4.28)$$

We are only interested in the solutions $w_k$ of eq.(4.26) with non-zero imaginary part and magnitude greater than 1. Taking the $n+1$'th root of the equation,

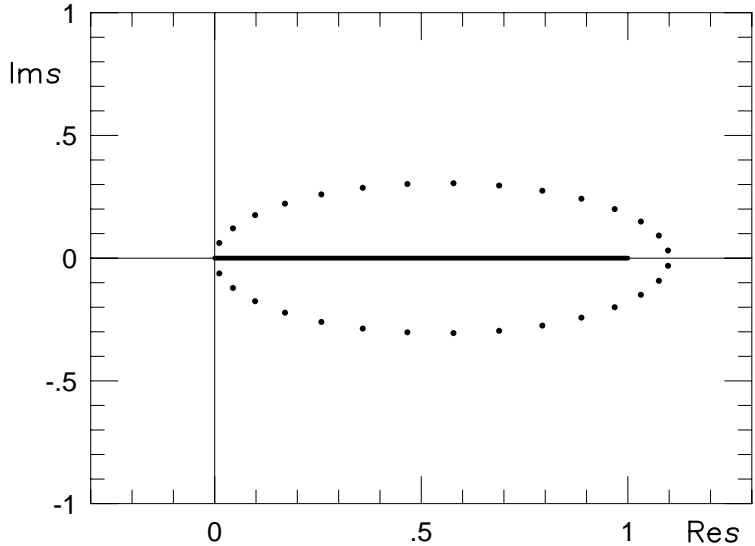

Fig. 1. Roots of $P(s)$ for $n = 30$ and $\varepsilon = 0.1$. The thick line represents the range $0 \leq s \leq 1$ of the spectrum of the quark matrix $Q^2$.

it is then straightforward to show that these are given by

$$w_k = -v_k \left\{1 - [(n+1)/(n-1)]v_k^{-2}\right\}^{-1/(n+1)} \left\{1 + \mathrm{O}(1/n^2)\right\}, \qquad (4.29)$$

where $v_k$ is defined through

$$v_k = r \exp\left\{i\frac{2\pi k}{n+1}\right\}, \qquad k = 1, \ldots, n. \qquad (4.30)$$

This result translates to an asymptotic expression for the roots $z_k$ through

$$z_k = \tfrac{1}{2}(1+\varepsilon) + \tfrac{1}{4}(1-\varepsilon)(w_k + 1/w_k). \qquad (4.31)$$

The approximation obtained in this way is quite accurate in most cases of interest.

For the numerical simulations one requires the roots to machine precision. This is now easy to achieve by applying a few iterations of Newton's algorithm to eq.(4.24), starting from the large $n$ values of the roots.

As shown by fig. 1, the roots tend to lie on an ellipse surrounding the spectral interval $0 \leq s \leq 1$. This pattern can be understood from the large $n$ expressions derived above. When only the leading terms are kept, the formulae reduce to

$$z_k = \tfrac{1}{2}(1+\varepsilon) - \tfrac{1}{2}(1-\varepsilon)\left\{\cosh\alpha\cos\frac{2\pi k}{n+1} + i\sinh\alpha\sin\frac{2\pi k}{n+1}\right\}. \qquad (4.32)$$

The smaller axis of the ellipse thus has a length equal to $\tfrac{1}{2}(1-\varepsilon)\sinh\alpha$. As $n$ increases the ellipse does not change very much, but the roots get denser with a separation between neighbours proportional to $1/n$.

*4.6 Summary*

The polynomial $P(s)$ defined in this section depends on the degree $n$ and a parameter $\varepsilon$. For fixed $\varepsilon$ and $n \to \infty$, it converges exponentially to $1/s$, for all spectral values $s$ in the range $0 < s \leq 1$. The rate of convergence is equal to

$$\alpha = 2\sqrt{\varepsilon} + \mathrm{O}(\varepsilon^{3/2}) \qquad (4.33)$$

if $s \geq \varepsilon$, and falls continuously to zero when $s < \varepsilon$.

We have also noted that the roots of $P(s)$ come in complex conjugate pairs with non-zero imaginary parts if $n$ is sufficiently large. The polynomial thus has all the properties required for the transformation to the bosonic theory to work out. It is a much better choice than the simple polynomial (3.1), because the rate of convergence at large $n$ is constant over an adjustable range of $s$. In particular, by tuning $\varepsilon$ we can avoid that most of the numerical effort goes into decreasing the approximation errors in places where they are already small.

We finally remark that the matrix inversion algorithm derived from this polynomial has an exponential rate of convergence close to $2/\sqrt{p}$ if we set $\varepsilon = 1/p$ (where $p$ is the condition number of the quark matrix).

## 5. Miscellaneous remarks

a. *Interpretation of the fields $\phi_k$.* The dominant contribution to the integral over $\phi_k$ in eq.(3.8) comes from a region in field space where $\|(Q - \mu_k)\phi_k\|$ is not much greater than $\nu_k\|\phi_k\|$. With high probability the field $\phi_k$ is hence found in the linear subspace spanned by all eigenstates of the quark matrix with eigenvalues in a range of width $\nu_k$ (or a few times $\nu_k$) around $\mu_k$. Note that $\nu_k$ is of order $\sqrt{\varepsilon}$ and so is much smaller than 1 in the cases of interest. $\phi_k$ may, therefore, be thought of as representing the contribution of the quark modes in a narrow spectral interval to the action of the effective gauge theory. Considering fig. 1 it is clear that the whole spectrum of the quark matrix is covered in this way.

b. *Choice of $n$ and $\varepsilon$.* As already noted before, the bosonic theory is expected to be accessible to standard numerical simulation techniques. In such calculations $n$ should be greater or equal to some minimal value $n_{\min}$ to guarantee that the systematic effects stemming from the finiteness of $n$ are negligible compared to the statistical errors. $n_{\min}$ depends on $\varepsilon$, the lattice parameters, the desired level of precision and the quantities to be computed. $\varepsilon$ should obviously be tuned so as to make $n_{\min}$ is as small as possible.

Extensive empirical tests will be needed to determine $n_{\min}$ and the optimal value of $\varepsilon$ in any given situation. We may, however, obtain some insight into the problem through the following qualitative argumentation.

It is quite clear that the spectral interval $\varepsilon \leq s \leq 1$, where uniform convergence is guaranteed, should contain most eigenvalues of the quark matrix $Q^2$. A crude free field estimate for the number $N_\varepsilon$ of levels with $s < \varepsilon$ is

$$N_\varepsilon = \frac{1}{32\pi^2} V M^4 \varepsilon^2, \tag{5.1}$$

where $V$ denotes the lattice volume and the quark mass has been set to zero for simplicity. On a $16^4$ lattice, for example, a value of $\varepsilon$ around 0.001 is required for at fixed physical conditions would grow $N_\varepsilon$ to be of order 1. If we then take $n = 380$ or so, the ratio $P'(s)/P(s)$ approximates $-1/s$ with an absolute accuracy better than $10^{-6}$ in the range $\varepsilon \leq s \leq 1$ [cf. eq.(4.23)].

These figures are perhaps a bit pessimistic and they should in any case not be taken too seriously. But they suggest that it may be necessary to set $n$ to values as large as a few hundred.

c. *Storage requirements.* On a $16^4$ lattice the memory size required to store 100 fields $\phi_k$ is about 157 Mwords. Although modern parallel computers tend to have a core memory sufficiently large to cope with such amounts of data, other ways of dealing with the storage problem exist.

On computers with fast I/O channels, for example, the field $\phi_k$ may be read in from an external storage device only when it is updated. This works out because $\phi_k$ does not couple to any field other than the gauge field (which is held in the main memory of the computer). Its contribution to the force acting on the gauge field can be computed at same time and may be added to the total force (which is also held in memory). After all $\phi_k$'s have been refreshed in this way, one may then proceed to update the gauge field. At this point there is no need to refer to the fields stored externally, because the force is already known.

d. *Critical slowing down and scaling.* The most efficient simulation algorithms for pure gauge theories are based on the idea of over-relaxation (for a recent review and references see ref.[7]). It is likely that similar methods apply to the bosonic theory discussed here. A further acceleration of the convergence of the algorithm may perhaps be achieved by adopting an update cycle, where the gauge field and the slow modes $\phi_k$ (those with a small value of $\nu_k$) are visited more often than the other modes.

If we assume that an algorithm with a dynamical critical exponent close to 1 can indeed be found, the computational effort at fixed physical conditions is expected to increase roughly like $a^{-6}$, when the lattice spacing $a$ is made smaller. At the same time the memory size required goes up approximately like $a^{-5}$. These estimates take into account that $\varepsilon$ must be scaled proportionally to $a^2$ to guarantee that most eigenvalues of the quark matrix remain in the spectral interval $\varepsilon \leq s \leq 1$. To preserve the level of accuracy it is then necessary to increase $n$ proportionally to $a^{-1}$.

It goes without saying that this argumentation is speculative and needs to be confirmed by empirical investigations.

e. *Correlation functions.* From a given ensemble of gauge field configurations, simulating the effective gauge field distribution $P_{\text{eff}}[U]$, the meson and baryon correlation functions can be obtained by computing the quark propagator using some iterative method such as the conjugate gradient algorithm. This may well be the best way to proceed, but it is nevertheless interesting to note that the correlation functions have an exact bosonic representation.

To deduce this representation we first observe that the correlation func-tions of interest can be rewritten as the expectation value of a sum of products of the bilocal operator

$$\mathcal{O}_{\alpha A,\beta B}(x,y) = \sum_{f=1,2} \overline{\psi}^{f}_{\alpha A}(x)\psi^{f}_{\beta B}(y) \tag{5.2}$$

($f$ denotes the flavour index, $\alpha,\beta$ the colour and $A,B$ the Dirac spinor indices). A generating functional for such expectation values is obtained by adding

$$a^8 \sum_{x,y} \sum_{\alpha,\beta} \sum_{A,B} J_{\alpha A,\beta B}(x,y)\mathcal{O}_{\alpha A,\beta B}(x,y) \tag{5.3}$$

to the QCD action (2.1), where $J$ is a suitable source field. The transformation to the bosonic theory now works out essentially as before, except that $Q$ is shifted by the source term. Subsequent differentiation with respect to the source then yields the desired bosonic representation of the hadron correlation functions.

It is worth pointing out that one ends up with correlation functions of local bosonic operators. The characteristic fermionic signs are taken into account when rewriting the product of the hadron operators in the form of sums of products of $\mathcal{O}_{\alpha A,\beta B}(x,y)$.

## 6. Conclusion

The local bosonic theory discussed in this paper provides a new starting point for numerical simulations of lattice QCD. Empirical studies are now needed to determine the computational cost of such simulations.

As a matter of principle one might object that the method involves an additional source of systematic error, because the number $n$ of bosonic fields coupled to the gauge field cannot be taken to infinity in practice. In many respects a finite value of $n$ amounts to an infrared cutoff in the quark sector and thus plays a rôle similar to the finite extent of the lattice. To be fair one should also say that all known simulation algorithms for dynamical fermions are affected with similar systematic errors. In the case of the Hybrid Monte Carlo algorithm, for example, one relies on an iterative scheme to compute the inverse of the quark matrix. Since only a finite number of iterations can be performed, one then needs to show that the ensuing error has a negligible effect on the simulation results.

I would like to thank Rainer Sommer, Peter Weisz and Ulli Wolff for discussions and critical comments on a first draft of this paper.